\documentclass[11pt]{article}\topmargin=-0.5cm\hfuzz=10pt  \oddsidemargin=-0.1cm \textheight 222mm\textwidth=16.3cm
\usepackage{amssymb}
\newcommand{\comm}[1]{}

\def\short{\comm}

\usepackage{color}

\newcommand{\bl}[1]{\color{blue}#1\color{black}}
\usepackage{endnotes}

\newtheorem{theorem}{Theorem}[section]
\newtheorem{proposition}[theorem]{Proposition}
\newtheorem{lemma}[theorem]{Lemma}
\newtheorem{corollary}[theorem]{Corollary}

\newtheorem{definition}[theorem]{Definition}
\newtheorem{remark}[theorem]{Remark}

\def\e{\varepsilon}

\def\defi{\stackrel{{\scriptscriptstyle \Delta}}{=}}
\def\defi{:=}

\def\a{\alpha}

\def\o{\omega}
\def\O{\Omega}

\def\F{{\cal F}}
\def\w{\widehat}
\def \Ind{{\,\rm Ind\,}}
\def \Ind{{\mathbb{I}}}

\def\esssup{\mathop{\rm ess\, sup}}

\def\R{{\bf R}}

\def\b{\beta}

\def\C{{\bf C}}

\def\ww{\widetilde}
\def\X{{\cal X}}
\def\t{\theta}
\def\oo{\bar}

\def\GG{{\cal G}}

\def\V{{\cal V}}
\def\A{{\cal A}}

\def\AA{{\cal C}}

\newcommand{\be}{\begin{equation}}
\newcommand{\ee}{\end{equation}}
\newcommand{\bd}{\begin{displaymath}}
\newcommand{\ed}{\end{displaymath}}
\newcommand{\ba}{\begin{array}{ll}}
\newcommand{\ea}{\end{array}}
\newcommand{\baa}{\begin{eqnarray}}
\newcommand{\eaa}{\end{eqnarray}}
\newcommand{\baaa}{\begin{eqnarray*}}
\newcommand{\eaaa}{\end{eqnarray*}}

\def\oo{\bar}

\def\a{\alpha}

\def\sinc{{\rm sinc}}

\def\ZZ{{\mathbb{Z}}}

\def\ee{\e}
\def\WO{\stackrel{p}{W_2^1}}

\date{Submitted:  May 16, 2024. Revised: June 18, 2024}

\title{Sampling Theorem and  interpolation formula   for non-vanishing signals
}
\author{Nikolai Dokuchaev}
 
\begin{document}
 \def\short{\comm}
\def\brea{}
\def\breakk{}
\def\break{}
\def\BRR{}
\def\breakm{\nonumber\\  }\def\BR{}\def\BRR{}
\def\breacm{}
\def\dt{}
\def\ZZM{\ZZ_{\scriptscriptstyle{\le 0}}}

\def\AA{{\cal A}}
\def\CC{{\cal C}}
\def\TP{{\frac{1}{2\pi}}}
\def\bl{}
\maketitle
 \begin{abstract} 
 The paper establishes an analog Whittaker-Shannon-Kotelnikov sampling theorem  with 
  with fast decreasing coefficient, as well as  a new modification of  the corresponding interpolation formula  applicable for general type non-vanishing bounded continuous signals.
 \end{abstract}


\section{Introduction}
Problems of recovery of signals from incomplete observations
 were studied intensively  in different settings. This includes 
 recovering signals from samples. 
The most important  tools  used for signal processing  are based on the representation of signal processes
in the frequency domain. This includes, in particular,   the conditions of  data recoverability. 
 In general, possibility  or recovery a   continuous time  signal from a sample is usually associated with restrictions on  the  class of underlying signals such as restrictions on the spectrum.
The classical Nyquist–Shannon sampling theorem  establishes that a band-limited signal  vanishing on $\pm \infty$ can be  recovered without error from a discrete sample taken with a sampling rate  that is at least twice the maximum frequency
of  the signal (the Nyquist critical rate). In particular, a band-limited  signal $x(t)\in C(\R)\cap L_2(\R)$ 
 with the spectrum contained in the interval $[-\pi,\pi]$ 
  can be recovered from its sample $\{x(k)\}_{k=-\infty}^\infty$ as
  \baa x(t)=\sum_{k=-\infty}^\infty \frac{\sin(\pi(k- t))}{\pi (k-t)}x(k).
  \label{Shan}\eaa 
 This is celebrated Whittaker-Shannon-Kotelnikov  interpolation formula, also known as 
 Whittaker–Shannon interpolation formula, Shannon's interpolation formula, and Whittaker's interpolation formula. It can be observed that since 
 the coefficients of this  interpolation formula are decreasing as $\sim 1/k$, it covers only signals  such that $x(t)\to 0$ with a certain rate  as $|t|\to +\infty$.

There are many works devoted to generalization of the Sampling Theorem; see e.g. the reviews in \cite{Eldar,KPT,Tr,V01,Z}  and literature therein. However, the extension of this result on non-vanishing signals has not been obtained yet. 

 The purpose of the present paper is to obtain an  interpolation formula similar to (\ref{Shan}) but applicable to non-vanishing continuous  signals.

 It can be observed that a non-vanishing  signal from $ L_{\infty}(\R) $ can be modified  to a signal from $L_1(\R)$ without any loss of information, for example, by replacement $x(t)$ by $e^{-|t|}x(t)$.  However,  at least for the case of  signals from $L_{1}(\R)$, these damping transformations represent  the convolutions in the frequency domain,  with smoothing kernels. Unfortunately, a band-limited signal will be transformed
 into a non-band-limited one along the way.  For the general type  two-sided processes  from $ L_{\infty}(\R) $, one could expect a similar impact of the damping transformations on the spectrum.
Therefore, it was essential to  develop a special  approach for extrapolation of non-vanishing signals from their samples.

The paper present an analog of Sampling Theorem and a modification of  Whittaker–Shannon-Kotelnikov interpolation formula   for non-vanishing signals (Theorem \ref{ThM} and formula (\ref{a}) in Section \ref{SecM}).  The $k$th coefficients for this  new interpolation formula (\ref{a})  are decreasing as $\sim 1/k^2$. 
Some numerical experiments are described in Section \ref{SecE}.

 \section{The main result: Sampling theorem and interpolation formula}
 \label{SecM}
 \subsection*{Some notations}

Let  $\R$ and $\C$, be the set of all real and complex numbers, respectively, and let $\ZZ$ be the set of all integers.

\def\DD{\mathbb{D}}

We denote by $ L_{\infty}(\R) $ the standard space  of all functions  $x:\R\to \C$,  considered up to equivalency,  such that
$\|x\|_{ L_{\infty}(\R) }\defi \esssup_{t\in\R}|x(t)|<+\infty$.

For $r\in[1,\infty)$, we denote by $L_r(\R)$  the standard space  of all functions  $x:\R\to \C$,  considered up to equivalency,  , such that
$\|x\|_{L_r(\R)}\defi \left(\int_{-\infty}^{\infty}|x(t)|^r dt\right)^{1/r}<+\infty$.

We denote by  $C( \R )$  the standard linear  space of continuous functions $f:  \R \to\C$
with the uniform norm $\|f\|_C\defi \sup_t |f(t)|$.

\begin{definition}\label{defG}
For a Borel measurable set  $D\subset  \R $ with non-empty interior,   let  
$x\in  L_{\infty}(\R) $ be such that  
$\int_{-\infty}^\infty x(t)y(t)dt=0$
for any  $y\in L_1(\R)$ such that  $Y|_{ \R \setminus D}\equiv 0$, where $Y$ is the Fourier transform of 
$y$. In this case, we say that $D$ is a spectral gap of $x\in  L_{\infty}(\R) $.
 \end{definition} 

For $\O\in (0,+\infty)$, we denote by  $\V(\O)$  the set of all signals $x\in L_\infty(\R)$ with
 the spectral gap $\R\setminus (-\O,\O)$. We call these signals band-limited.

We use the terms "spectral gap" and "band-limited" because, for  a signal  $x\in L_2(\R)$, Definition \ref{defG} means that $X(\o)=0$ for $\o\in D$, where $X$ the Fourier transform of $x$.
 The standard Fourier transform is not applicable for general type non-vanishing signals from $L_\infty(\R$, however, we will use the terms "spectral gap" and "band-limited" for them as well. It is shown in Section \ref{ssecBL} below that this is still justified with respect to the spectral properties of these signal.

 Let $\O\in (\O,\pi)$ be given.
Let some  even integer number $N$ be selected such that \baaa\label{N}
N>\frac{\O}{\pi-\O}.
\eaaa
Further, let some $\O_1\in (\O,\pi)$ be selected such that \baaa\label{N1}
N\ge \frac{\O_1}{\pi-\O_1}.
\eaaa
Clearly, such $\O_1$  exists.
 
For $t\in [N,N+1)$ and  $\tau=t-N\in[0,1)$,  let us select \baaa
g(t)=\frac{\pi \lfloor t\rfloor }{t}=\frac{\pi N}{N+\tau},
\eaaa
\bl{where $\lfloor t\rfloor\defi \min\{n\in\ZZ:\ t\le n\}$. }   It is easy to see that,  for any  $t\in [N,N+1)$, we have  $g(t)\ge \pi N(N+1)^{-1}\ge \O_1$ and $g(t)t=\pi N$.

Assume that the function $g(t)$ is extended periodically from $[N,N+1)$ to $g:\R\to [\O_1,\pi]$.  
This function is right-continuous. In addition,  $g(m)=\pi$ and $(t-m)g(t)=\pi N$   for any integer $m$ and any $t\in[N+m,N+m+1)$. 
\begin{theorem}\label{ThM}  For any continuous bounded  band-limited signal  $x\in C(\R)\cap \V(\O)$,
for any integer $m$ and any $t\in[N+m,N+m+1)$, we have that
\baa
x(t)=\sum_{k\in\ZZ }a_k(t)x(k),
\label{a0}\eaa
where $a_m(t)=1-\frac{g(t)}{\pi}$, and 
\comm{Atau[k]=N*sinc(gg*(ttD[k]-mmm))/((ttD[k]-tau1))}
\baa
a_k(t)=\frac{(t-m)\sin[ g(t)(k-m)]}{\pi (k-m)(k-t)},\qquad k\neq m. \label{a}
\eaa 
 The corresponding series is absolutely convergent. 
 \end{theorem}
In particular, formula (\ref{a}) implies that $a_k(k)=1$, and that $a_k(l)=0$ for any integers $k$ and $l$ such that $k\neq l$.

It can be noted that, under the assumptions of Theorem  \ref{ThM},  we have that    
\baaa
a_k(t)
=\frac{(t-m)g(t)\sinc[ g(t)(k-m)]}{\pi(k-t)}
=\frac{N\sinc[ g(t)(k-m)]}{ k-t},\quad k\neq m. \label{alt2}
\eaaa
We used here that  $(t-m)g(t)=\pi N$.
\begin{corollary}\label{corrS1}  Theorem \ref{ThM} implies that a   non-vanishing process from $x\in \V(\R)\cap C(\R)$, i.e., a band-limited process with  the spectral gap
 $\R\setminus (-\O,\O)$,   is uniquely defined for any $\t\in\R$ by its one-sided sample $\{x(k)\}_{k\in\ZZ,k\le \t}$. 
 \end{corollary}
\section{Some numerical experiments}\label{SecE}
In some straightforward numerical experiments, 
we applied truncated interpolation classical interpolation formula 
(\ref{Shan}) and our formula (\ref{a}) for simulated band limited signals, with summation over $\{k\in\ZZ:\ |k|\le L\}$, for some large enough $L$. For these experiments, we used 
$\O=\frac{5}{12}\pi$ and preselected $\O_1=(\O+\pi)/2$. 
The number  $N$ was selected as the smallest   even number such that $N>\O_1/(\pi-\O_1)$.
It can be noted that these choices define the simulated signal uniquely.
We experimented with band-limited vanishing signals from $L_2(\R)$ as well as with non-vanishing signals
from $L_\infty(\R)$.

For band-limited vanishing signals from $L_2(\R)$, we found that the results were
 indistinguishable for  both formulae (\ref{Shan}) and (\ref{a}).   In particular, we considered a  band-limited signal 
$x(t)=A[\sinc(M\pi t)+\sinc(M\pi (t-1)/2]$, with $M=256$ and $A=\sqrt{M 4/5}$. 
 This signal has been used  for numerical examples in \cite{KPT}, p.30. 

We estimated $x(t)$ at several arbitrarily selected single points. 
For example, for $t=47830.4$, we  found the following.
\begin{itemize}
\item
The  error for both interpolation  for $L=10^3$ was about $10^{-5}$.
\item
The  error for both interpolation  for $L=10^4$ was about $10^{-6}$.
\item
The  error for both interpolation for $L=10^5$ was about 
$10^{-7}$. More precisely,  the error for interpolation  (\ref{Shan}) 
 was $1.3503206299415515963\cdot 10^{-7}$,  and the error for interpolation  (\ref{a}) 
 was $7.6562947048617628244\cdot 10^{-7}$.
 \end{itemize}

In addition, we tested these interpolation formulae for some non-vanishing processes.
 In particular, we considered signal  
 $x(t)=\cos(\O t-L/2))$. Again, we estimated $x(t)$ at several arbitrarily selected single points. 
For  $t=47830.4$, we  found the following.
\begin{itemize}
\item
For $L=10^3$, the  errors for the interpolation was about  $10^{-5}$ 
 for   (\ref{Shan}), and about  $10^{-6}$ 
 for   (\ref{a}).
\item
For $L=10^3$, the  errors for the interpolation was about  $10^{-5}$ 
 for   (\ref{Shan}), and about  $10^{-6}$ 
 for   (\ref{a}).
\item
For $L=10^4$, the  errors for the interpolation was about  $10^{-5}$ 
 for   (\ref{Shan}), and about  $10^{-8}$ 
 for   (\ref{a}).
\item
For $L=10^5$, the  errors for the interpolation was about  $10^{-6}$ 
 for   (\ref{Shan}), and about  $10^{-11}$ 
 for   (\ref{a}). More precisely, the error for interpolation  (\ref{Shan}) 
 was $1.577106079952983464\cdot 10^{-6}$,  and the error for interpolation  (\ref{a}) 
 was $ 8.7279850013999293878\cdot 10^{-11}$.
 \end{itemize}
These experiments  show that formula (\ref{a})  can replace  formula (\ref{Shan}) for band-limited  vanishing signals, 
and can be used for effectively  for non-vanishing signals as well. 

 \section{Background:  spectral representation for non-vanishing signals }
 \label{SecB}
 In this section, we outline  some results being used  in the proof for the main Theorem \ref{ThM}.
\subsection{Some notations and  definitions for spaces of functions}
We denote by  $\overline{z}$ the complex conjugation. We denote by $\ast$ the convolution 
\baaa
(h\ast x)(t)\defi\int_{-\infty}^\infty h(t-s) x(s)ds, \quad t\in\R.
\eaaa

For a Banach space $\X$, we denote by $\X^*$ its dual.

For $r\in[1,\infty)$, we denote by $\ell_r$ the set of all processes (signals) $x:\ZZ\to \C$, such that
$\|x\|_{\ell_r}\defi \left(\sum_{t=-\infty}^{\infty}|x(t)|^r\right)^{1/r}<+\infty$. We denote by $\ell_\infty$ the set of all processes (signals) $x:\ZZ\to \C$, such that
$\|x\|_{\ell_\infty}\defi \sup_{t\in\ZZ}|x(t)|<+\infty$.

We denote by   $ W_2^1(\R)$  the Sobolev  space of functions $f:  \R \to\C$
that belong to $L_2\R $ together with the distributional derivatives
up to the first order.

Clearly, the embeddings 
$ W_2^1 (\R)\subset C( \R )  $ and 
$  C(\R)^*\subset W_2^1 (\R)^*$ are continuous.

Let  ${\WO}(-\pi,\pi)$ 
denote  the Sobolev  space of functions $f: [-\pi,\pi]\to\C$
that belong to $L_2(-\pi,\pi)$ together with the distributional derivatives
up to the first order, and such that $f(-\pi)=f(\pi)$.

Let  $\CC$  be  the space of functions $f\in C([-\pi,\pi])$
with the  finite norm $\|f\|_{\CC}\defi \sum_{k\in\ZZ}|\w f_k|$, where 
$\w f_k=\frac{1}{2\pi}\int_{-\pi}^\pi e^{-i\o s}f(s)ds$ are the Fourier coefficients of $f$.
In other words, $\CC$ is the space of absolutely convergent Fourier series on $[-\pi,\pi]$. 
By the choice of its norm, this is a separable Banach space that is isomorphic to $\ell_1$.

\begin{lemma}\label{lemma1} 
\begin{enumerate}
\item
 The embedding 
$\WO(-\pi,\pi)\subset\CC$ is continuous.
\item
If $f\in \CC$ and $g\in\CC$, then $h=fg\in\CC$, and $\|h\|_{\CC}\le \|f\|_{\CC}\|g\|_{\CC}$.   
\item For $f\in \CC$ , define  $g_f(\o,m)\defi e^{i m\o}f(\o)$, where $m\in\ZZ$, $\o\in\R$. Then $g_f(\cdot,t)\in\CC$ and 
$\|g_f(\cdot,t\|_{\CC}=\|f\|_{\CC}$.
\end{enumerate}
\end{lemma}

Let $\AA$  be the space of continuous functions $f\in C( \R )$
with the  finite norm $\|f\|_{\AA}\defi \int_{\R}|\w f(\o)|d\o$, where 
$\w f(\o)=\int_{-\infty}^\infty e^{-i\o s}f(s)ds$ is the Fourier transform of $f$.
By the choice of this norm, this is a separable Banach space that is isomorphic to $L_1(\R)$. 

In particular, the definition for $\AA$ implies that $Y\in\AA$ in Definition \ref{defG}.

It can be noted that there are functions  in $C( \R )$ that do not belong to $\AA$.

\begin{lemma}\label{lemma2} 
\begin{enumerate}
\item The embedding 
$ W_2^1 (\R)\subset\AA$ is continuous.
\item
If $f\in \AA$ and $g\in\AA$, then $h=fg\in\AA$, and $\|h\|_{\AA}\le \|f\|_{\AA}\|g\|_{\AA}$.   
\item For $f\in \AA$ , define  $g_f(\o,t)\defi e^{i t\o}f(\o)$, where $t,\o\in\R$. Then $g_f(\cdot,t)\in\AA$,
$\|g_f(\cdot,t\|_{\AA}=\|f\|_{\AA}$, and the function  $g(\cdot,t)$ is continuous in $\AA$ with respect to $t\in\R$.
\end{enumerate}
\end{lemma}

In particular, it follows that  the embeddings 
$
W_2^1 (\R) \subset\AA \subset C( \R ) \subset L_1( \R )$ and 
$ L_1( \R )^*\subset  C( \R )^* \subset \AA^*\subset  W_2^1 (\R)^*$
are continuous. 

We assume that each $X\in L_1( \R )$ represents an element of the dual space  $C( \R )^*$ such that
 $\langle X,f\rangle=\int_{-\infty}^\infty X(\o)f(\o)d\o$ for $f\in C( \R )$. We will use the same notation
  $\langle \cdot,\cdot\rangle$ for the extension of this bilinear form  
  on  $\AA^*\times \AA$.

\subsection{Spectral representation for non-vanishing signals }
\label{ssecSpec}
The space $\AA$ and its dual $\AA^*$ have been used  to define formally a  spectral representation 
for $x\in \ell_\infty$ via $X\in\AA^*$  
such that 
$\langle X,f\rangle=\int_{-\infty}^\infty x(t)\varphi(t)dt$ for any $f\in \AA$, where $\varphi\in L_1(\R)$  is  the Fourier transfer for $f$; see, e.g., Chapter VI in \cite{Katz}.  
In Chapter III in \cite{Kahane}, a similar definition was used for the Fourier transforms for 
pseudo-measures on $[-\pi,\pi]$ represented as elements of $\ell_\infty$. 
  However, for the  purposes
of this paper,   we will use a more straightforward  definition 
from \cite{D24} based on the following lemma. 

\begin{proposition}\label{Prop1}  For any $x\in  L_{\infty}(\R) $, there exists a 
weak* limit   $X\in \AA^*$ of the sequence of functions 
$X_m(\o)\defi \int_{-m}^m e^{-i\o t} x(t)dt$ defined on $ \R $ for $m>0$.
This $X$ is such that $\|X\|_{\AA^*}=\|x\|_{L_\infty(\R)}$.    
\end{proposition}

 It can be noted that, in Proposition \ref{Prop1},  $X_m\in L_1( \R ) \subset C( \R )^*\subset \AA^*$.

 We define a spectral representation of $x\in L_{\infty}(\R)$ via   mapping  $\F: L_{\infty}(\R) \to \AA^*$  such that  $X=\F x$ for $x\in  L_{\infty}(\R) $ is  the  limit  in $\AA^*$  introduced in Proposition \ref{Prop1}. By  Proposition \ref{Prop1}, this mapping is linear and continuous.
 
 \def\GGG{{\cal M}}

 Clearly, for $x\in L_1(\R)$, $\F x$ is the standard Fourier transform, and $\GG=\F^{-1}$ is 
 the inverse Fourier transform.
 
 Further, for any $h\in L_1(\R)$, define a mapping  $\GGG_h:\AA^*\to  L_{\infty}(\R) $ such that
$y_h=\GGG_h X$  is defined as \baaa
y_h(t)=\TP\langle X, H(\cdot)e^{i\cdot t} \rangle \quad \hbox{for}\quad X\in \AA^*, \quad H=\F h,\quad t\in\R.
\eaaa
By Lemma \ref{lemma1}(iii), it follows    that $H(\cdot)e^{i\cdot t}\in \AA$ for any $t\in\R$, and 
it is continuous in $t$ in the topology of $\AA$. 

\begin{remark} For the special case where  $X\in L_1(\R)$, the standard results for Fourier transformations imply  for   $h\in\AA$ that $y_h(t)=\TP\langle Y_h,e^{i\cdot t}\rangle$ for any $t\in\R$,
where 
 $Y_h =H X$.   In this case, the form $\langle H X,e^{i\cdot t}\rangle$ is well defined since 
 $H\in C(\R)$ and hence
 $H X\in L_1(\R)$. 
 \end{remark}

Clearly, the operator $\GGG_h: \AA^*\to  L_{\infty}(\R) $ is linear and continuous for any $h\in\AA$. Moreover,   $y_h(t)$ is continuous in $t$,  $\GGG_h(\AA^*)\subset  C(\R)$, and  the mapping 
$\GGG_h: \AA^*\to  C(\R)$ is continuous.

\begin{lemma}\label{lemmahxy}
 \begin{enumerate}
 \item For any $x\in L_\infty(\R)$ and $X=\F x$, we have that $(h\ast x) (t)=y_h(t)$, where  $y_h=\GGG_h X$.
 \item For any $X\in\AA^*$ and  $y=\GGG_h X$, there exists an unique up to equivalency process 
 $x\in L_{\infty}(\R)$ such that $(h\ast x)(t)=y_h(t)$ for any $h\in \AA$ for all $t$. For this process, we have that $\|x\|_{L_{\infty}(\R)}\le \|X\|_{\AA^*}$,
 and $\F x=X$.
 \end{enumerate} 
\end{lemma}

We define an operator $\GG:\AA^*\to  L_{\infty}(\R)$ such that $x=\GG X$ in Lemma \ref{lemmahxy}(ii) above.

\begin{theorem}\label{Th1} The mappings  $\F: L_{\infty}(\R) \to\AA^*$ and 
$\GG:\AA^*\to  L_{\infty}(\R) $ are continuous isometric bijections such that 
$\F=\GG^{-1}$ and  $\GG=\F^{-1}$. 
\end{theorem}

 \subsection{Band-limited signals and spectral representation}
 \label{ssecBL}
   
The following lemma connects Definition \ref{defG} with the spectral representation. 
\begin{lemma}\label{lemmaG} 
A signal  $x\in L_\infty(\R)$ has a spectral gap $D\subset \R$ if and only if  
$\langle  \F x,f\rangle =0$ for any $f\in \AA$ such  that $f|_{ \R \setminus D}\equiv 0$.
\end{lemma}

 This implies that, for any signal $x\in \V(\O)$ and any $f_1,f_2\in\AA$,
  if $f_1(\o)=f_2(\o)$ for all $\o\in[-\O,\O]$  then $\langle  \F x,f_1\rangle=\langle  \F x,f_2\rangle$.


\section{Proofs}
\subsection{Proofs of auxiliary results}
 The proof for Lemma \ref{lemma1} can be found in \cite{D23}.

{\em Proof of Lemma  \ref{lemmaG}}. Let  $y\in L_1(\R)$, and let  $Y=\F y\in L_\infty(\R)$.
 It follows from  the definitions  that $Y\in\AA$.  Let $h(t)=\oo y(-t)$ and $H= \F h$. 
 We have that $Y=\oo H$.
 Furthermore, \baaa  \int_{-\infty}^\infty x(t)y(t)dt=\int_{-\infty}^\infty x(t)\oo h(-t)dt=(\oo h\ast x)(0)=\TP\langle X, H e^{i\cdot 0} \rangle
 =\TP\langle X, H \rangle =\TP\langle X,\oo Y \rangle.
\eaaa
Since $Y(\o)=0$ if and only if $\oo Y(\o)=0$, the lemma statement follows from the definitions. 
 $\Box$
 
The proofs for the remaining  statements listed 
in Section \ref{ssecSpec}  can be found in \cite{D24}.

\subsection{Proof of Theorem \ref{ThM}}
 As the first step to prove  Theorem \ref{ThM}, we  need  to obtain its more abstract  conditional version.
 \def\aa{{\rm{a}}}
\def\ab{b}

\begin{proposition}\label{lemmaE}  Let $\O\in (0,\pi)$ be given, and let $\O_1\in (\O,\pi)$ be selected.
 Suppose that there exists a function $E:\R\times\R\to \C$ such that the following holds. 
\begin{enumerate}
\item $E(t,\o)=e^{i \o t}$ for all $t\in\R$ and $\o\in [-\O_1,\bl{\O_1}]$.
\item
For any $t$,  $E(t,\cdot)|_{[-\pi,\pi]}\in\CC$ and $ \sup_{t\in\R}\|E(t,\cdot)|_{[-\pi,\pi]}\|_\CC<+\infty$.
\item
For any $t$,  $E(t,\cdot)\in\AA$ and $ \sup_{t\in\R}\|E(t,\cdot)\|_\AA<+\infty$.
\end{enumerate}
For $t\in\R$ and $k\in\ZZ$, let 
\baaa
\aa_k(t)\defi \frac{1}{2\pi}\int_{-\pi}^\pi E(t,\o)e^{-i\o k}d\o. \eaaa
Then any signal $x\in C(\R)\cap \V(\O)$ can be represented as
\baaa
x(t)=\sum_{k\in\ZZ }\aa_k(t)x(k).
\label{a00}
\eaaa
The corresponding series is absolutely convergent. In addition, if  $E(t,\o)=\overline{E(t,-\o)}$  for all  $t$ and $\o$,  then $\aa_k(t)\in\R$.
\end{proposition}

\par
It can be noted that, under the assumption of Lemma \ref{lemmaE},
  we have that 
 \begin{enumerate}
 \item $\aa_k(k)=1$ and $\aa_k(m)=0$ for all $t\in\R$ and all integers $k$ and $m$, $m\neq k$;
\item $\{\aa_k(t)\}_{k\in\ZZ}\in\ell_1$ for all $t$.
  \end{enumerate}

{\em Proof of Proposition \ref{lemmaE}}.
Suppose that  $E(t,\o)$ is  such as described in Lemma \ref{lemmaE}.
Since $E(t,\cdot)\in\AA$, we have that $\{\a_k(t)\}\in\ell_1$ for all $t$. Hence   \baaa
E(t,\o)=\sum_{k\in\ZZ} \aa_k(t) e^{i \o k}=\sum_{k\in\ZZ} \aa_k(t) E(k,\o),\quad t\in \R,\quad \o\in
[-\O_1,\O_1],
\eaaa 
where the series are absolutely convergent for any $t$, $\o$. Moreover, the sum
\baa \label{EE}
E(t,\cdot)=\sum_{k\in\ZZ} \aa_k(t) E(k,\cdot)
\eaa   converges in $\A$ for any $t\in \R$.

It can be reminded  that
$e^{i\o t}=\sum_{k\in\ZZ} \aa_k(t) e^{i \o k}$ for $\o\in[-\O_1,\O_1]$, but this does not hold if $|\o|>\O_1$.  

Further, let $\AA_\O$ be the set of all $h\in\AA$ such that $h(t)=0$ if $|t|>\pi-\O_1$. Clearly, $(h\ast e^{i\cdot t})(\o)=(h\ast E(t,\cdot))(\o)$ if $\o\in[-\O,\O]$ and $h\in\AA_\O$.  Let $X=\F x$. By Lemma \ref{lemmahxy}, by Theorem \ref{Th1}, and by  the choice of $E$ and $X$, we have that \baaa
y_h(t)=(h\ast x)(t)=\frac{1}{2\pi} \langle X, h\ast e^{i\cdot t} \rangle= \frac{1}{2\pi} \langle X, h\ast E(t,\cdot ) \rangle  \eaaa
for all $t$ and all $h\in\AA_\O$. Hence
\baaa
x(t)=   \frac{1}{2\pi}\langle X, E(t,\cdot )  \rangle ,\quad t\in\R. \eaaa
In particular, \baaa
x(k)= \frac{1}{2\pi}\langle X, E(k,\cdot ) \rangle,\quad k\in\ZZ.
\eaaa
By (\ref{EE}), it follows that
\baaa
x(t)=   \frac{1}{2\pi}\langle X, \sum_{k\in\ZZ} \aa_k(t) E(k,\o)\rangle =\sum_{k\in\ZZ} \aa_k(t)\frac{1}{2\pi}\langle X,  E(k,\o)\rangle  
= \sum_{k\in\ZZ} \aa_k(t)x(k).\eaaa
This completes the proof of Lemma \ref{lemmaE}.
$\Box$

The following step is to find a function $E$ satisfying the assumptions of Proposiotion \ref{lemmaE}.

Up to the end of this paper, we assume that $N$ and $g$  are selected as in Theorem \ref{ThM}.
In particular,  $g(m)=\pi$ for any $m\in\ZZ$.

\begin{lemma}\label{lemmaNg}  We have that $e^{ig(t)t}=e^{-ig(t)t}=1$ for 
all $t\in [N,N+1)$. In addition,  $e^{ig(t)(t-m)}=e^{-ig(t)(t-m)}=1$ for 
all  $m\in\ZZ$ and $t\in [N+m,N+m+1)$.
\end{lemma}

{\em Proof of  Lemma \ref{lemmaNg}}. 
Let $t=N+\tau$ and $\tau\in[0,1)$. We have \baaa
g(t)=\frac{\pi \lfloor t\rfloor }{t}=\frac{\pi N}{N+\tau}.
\eaaa
Clearly, $\O_1\le g(t)\le \pi$ , $g(t)t=\pi N$ and  
\baaa
e^{i g(t) t}=e^{i \pi N}=1, \quad t\in[N,N+1).
\eaaa
We used here that $N$ is  even. Further, let $m\in\ZZ$. By the choice of $g$, we have that $g(t+m)=g(t)$, and 
\baaa
e^{i g(t) (t-m)}=e^{i g(t-m) (t-m)}=1, \quad t\in[N+m,N+m+1).
\eaaa
This completes the proof of Lemma \ref{lemmaNg}. $\Box$

\begin{lemma}\label{lemmaE2}
 Let  a function  $\ww E:[N,N+1\bl{)\times\R}\to \C$ be defined  
as 
\baaa
&\ww E(t,\o)=e^{i \o t},\quad &\o\in [-g(t),g(t)],\\
&\ww E(t,\o)=e^{i g(t)t}, \quad         &\o\notin [-g(t),g(t)].
\eaaa Further, let  a function $\xi:[N,N+\bl{1)}\times\R\to\R$ be selected such that $\xi(t,\cdot)\in W_2^1(\R)$ for any $t$, \bl{ $\xi(t,\o)=\xi(t,-\o)$ for any $t$ and $\o$, }
and  $\xi(t,\o)=1$ for any $\o\in\bl{[-\pi,\pi]}$. We define the function $E_N:\bl{[N,N+1)}\times \R\to\C$ as 
 \baaa
E_N(t,\o)=\ww E(t,\o)\xi(t,\o), \quad &\o\in \R, \quad t\in[N,N+1\bl{)}.
\eaaa
Further, let  $t\in[N+m,N+m+1)$, where  $m\in\ZZ$.
Let $\tau=t-m$; we have that $\tau \in[N,N+1)$. Let 
\baaa
E(t,\o)=E_N(\tau,\o) e^{i\o m}, \quad &\o\in \R,\quad t\in[N+m,N+m+1),\quad \tau=t-m. 
\eaaa
Then the conditions (i)-(iii) of  Proposition \ref{lemmaE} hold for these  $E$.  In addition, $E(t,\o)=\overline{E(t,-\o)}$  for all  $t$ and $\o$.
\end{lemma}
\newpage
\par
{\em Proof of  Lemma \ref{lemmaE2}}.  It is easy to see that condition (i) of Proposition \ref{lemmaE} is satisfied for $E$. 

Further, 
$\bl{E_N(t,\cdot)|_{[-\pi,\pi]}=}\ww E(t,\cdot)|_{[-\pi,\pi]}\in \WO(-\pi,\pi)$ for any $t\in[N,N+1)$, and \baaa
\sup_{t\in[N,N+1\bl{)}}\|\bl{E_N}(t,\cdot)|_{[-\pi,\pi]}\|_{\WO(-\pi,\pi)}<+\infty.
\eaaa Hence $E_N(t,\cdot)|_{[-\pi,\pi]}\in\CC$  for any $t\in[N,N+1)$ and 
\baaa
\sup_{t\in[N,N+1\bl{)}}\|\bl{E_N(t,}\cdot)|_{[-\pi,\pi]}\|_{\CC}<+\infty.\eaaa 

Furthermore, 
$E_N(t,\cdot)\in W_2^1(\R)$ for any $t\in[N,N+1)$, and \baaa
\sup_{t\in[N,N+1\bl{)}}\|\bl{ E_N}(t,\cdot)\|_{W_2^1(\R)}<+\infty.\eaaa 
 Hence $E_N(t,\cdot)|_{[-\pi,\pi]}\in\AA$  for any $t\in[N,N+1)$ and 
\baaa
\sup_{t\in[N,N+1\bl{)}}\|\bl{E_N(t,}\cdot)|_{[-\pi,\pi]}\|_{\bl{\AA}}<+\infty.\eaaa

Hence  conditions (i)-(ii) of Proposition \ref{lemmaE} are satisfied  for $\bl{E_N}(t,\o)$ for $t\in[N,N+1)$. 

Further, by Lemma \ref{lemma1}(iii), for any 
$v\in\AA$, we  have that \baaa  e^{i\cdot m}v\in \AA,\quad \|v\|_{\AA}=\|e^{i\cdot m}u\|_{\AA}.\eaaa
Similarly, we have for $u\in\CC$  that \baaa
e^{i\cdot m}u\in \CC,\quad 
 \|u\|_{\CC}=\|e^{i\cdot m}u\|_{\CC}.\eaaa Then conditions (iii) are satisfied  for the selected  $E$. \bl{
 Hence }  condition (iii) of Proposition \ref{lemmaE} is satisfied  for \bl{$E(t,\o)$. }
 In addition,   $E(t,\o)=\overline{E(t,-\o)}$  for all  $t$ and $\o$.
This completes the proof of Lemma \ref{lemmaE2}.
$\Box$

It can  be noted that $E(t,g(t))=\oo E(t,-g(t))\neq E(t,-g(t))$ for $E$ selected as in  Lemma \ref{lemmaE2}
for non-integer $t\not\in\{N,N+1\}$.

We will denote by $a_k(t)$ the corresponding coefficients $\aa_k(t)$ defined as in Proposition \ref{lemmaE}
with $E$ and $g$ defined by  Lemma \ref{lemmaE2}.

 \begin{proposition}\label{propS1} 
 Let  $E$ be selected as in Lemma \ref{lemmaE2}. For $t\in[N,N+1)$, we have that $a_0(t)=1-\frac{g(t)}{\pi}$ and
\baaa
a_k(t)=\frac{t\sin[ g(t)k]}{\pi k(k-t)},\quad k\neq 0.\label{a000}
\eaaa
 \end{proposition}
 
It  can be noted that, since $g(t)t=\pi N$ for $t\in[N,N+1)$, we have that  \baaa
a_k(t)=\frac{g(t) t\, \sinc(g(t)k)}{\pi(k-t)}=\frac{N\sinc(g(t)k)}{k-t},\quad k\neq 0. \label{alt}
\eaaa

\par
{\em Proof of Proposition \ref{propS1}}.  Clearly,  $a_k(t)=\Ind_{t=k}$ for $t\in\ZZ$, by 
the choice of $g(k)=\pi$. Further,  we have that
\baaa
a_k(t)=\frac{1}{2\pi}(\a_k(t)+\b_k(t)), \label{ak1}
\eaaa
where 
 \baaa
&&\a_k(t)=\int_{-g(t)}^{g(t)} e^{-i\o k}  e^{i\o t} d\o,\qquad
\\ &&\b_k(t)=\int_{-\pi}^{-g(t)} e^{-i\o k} e^{i g(t)t} d\o
+\int_{g(t)}^{\pi} e^{-i\o k}  e^{i g(t)t} d\o=e^{i g(t)t}
\left(\int_{-\pi}^{-g(t)} e^{-i\o k} d\o
+\int_{g(t)}^{\pi} e^{-i\o k} d\o\right). \label{ab1}
\eaaa
Assume that $k\neq 0$. In this case, 
 \baaa
\a_k(t)=\int_{-g(t)}^{g(t)} e^{-i\o k}  e^{i\o t} d\o=\frac{e^{ig(t)(t-k)}-e^{-ig(t)(t-k)} }{i(t-k)}=
\frac{e^{-ig(t)k}-e^{ig(t)k} }{i(t-k)}
=-\frac{2\sin(g(t)k)}{t-k},\label{ab}
\eaaa
\baaa
\b_k(t)=e^{i g(t)t}
\left(\frac{e^{ig(t)k}-e^{i\pi k} }{-ik}+\frac{e^{-i\pi k}-e^{-ig(t)k} }{-ik}\right)=e^{i g(t)t}
\left(\frac{e^{-ig(t)k} -e^{ig(t) k} }{ik}\right)\\=-e^{i g(t)t}\frac{2\sin(g(t)k)}{k},
\eaaa
and
\baaa
a_k(t)=\frac{1}{2\pi}(\a_k(t)+\b_k(t)) =-\frac{1}{2\pi}\left(\frac{2\sin(g(t)k)}{t-k} + e^{i g(t)t}\frac{2\sin(g(t)k)}{k}\right).  \label{a2}
\eaaa
By the choice of even $N$, we have that $ e^{i g(t)t}= e^{i \pi N}=1$. Hence
\baa
a_k(t)=\frac{1}{2\pi}(\a_k(t)+\b_k(t)) =
-\frac{1}{\pi} \sin(g(t)k)\left(\frac{1}{t-k} +\frac{1}{k}\right) 
=
\frac{1}{\pi} \sin(g(t)k)\frac{t}{k(k-t)}.
 \label{a1}
\eaa
\par
Further, assume that $k=0$. In this case,  
\baaa
\a_0(t)=\int_{-g(t)}^{g(t)}   e^{i\o t} d\o=-\frac{e^{ig(t)t}-e^{-ig(t)t} }{i t }=
-\frac{e^{iN\pi}-e^{-iN\pi} }{i t }. \label{ab2}
\eaaa
Since $t\ge N>0$, we have that $\a_0(t)=0$. Further, we have
  $\b_0(t)=2e^{i g(t)t}(\pi-g(t))=2(\pi-g(t))$. Hence  $a_0(t)=1-\frac{g(t)}{\pi}$.
 This completes the proof of Proposition \ref{propS1}. $\Box$

\begin{lemma}\label{lemmaShift} Let  $E$ be selected as in Lemma \ref{lemmaE2}. 
 For any  $k,m\in\ZZ$, we have that 
 \baaa
a_k(t+m)=a_{k-m}(t).
 \eaaa 
 \end{lemma}
 \par
{\em Proof of Lemma \ref{lemmaShift}}. Let $N$  be defined as  in Lemma \ref{lemmaE2},
and let $t=N+\tau$, 
 By the definitions, $E(t,\o)=e^{iM\o}E(\tau,t)$, where $M\in \ZZ$ is such that $\tau=t-M\in [N,N+1)$.
 Hence $E(t+m,\o)=e^{im\o}E(t,\o)$ and
 \baaa
a_k(t+m)=\frac{1}{2\pi}\int_{-\pi}^\pi e^{im\o}E(t,\o)e^{-i\o k}d\o=
\frac{1}{2\pi}\int_{-\pi}^\pi E(t,\o)e^{-i\o (k-m)}d\o=a_{k-m}(t). 
 \eaaa  This completes the proof of Lemma \ref{lemmaShift}.
$\Box$
\begin{corollary}\label{corrShift} Let  $E$ be selected as in Lemma \ref{lemmaE2}.
Let  $a_k(\cdot)$ be defined by (\ref{a}). Then for any $m\in\ZZ$, any signal $x\in \V(\O)\cap C(R)$ can be represented, for $t\in[N+m,N+m+1)$, as
\baa
x(t)=\sum_{k\in\ZZ }a_{k-m}(t-m)x(k).
\eaa
The corresponding series is absolutely convergent. 
\end{corollary}
\begin{remark} Corollary \ref{corrShift} is
due to the particular choice of acceptable $E$. 
Possibly,  there exist  acceptable choices of $E$ such that  does not hold. 
\end{remark}

 {\em Proof of Corollary  \ref{corrShift}}.  Let $\ww x(t)\defi x(t+m)$. It is easy to see that $\ww x\in\V(\O)$. Clearly, $x(t)=\ww x(t-m)$ for all $t$. By Proposition \ref{propS1}, we have that
  \baaa
\ww x(s)=\sum_{k\in\ZZ } a_{k}(s)\ww x(k),\quad s\in [0,1).
\eaaa
Hence, for $t\in [m+N,m+N+1)$,
\baaa
x(t)=\ww x(t-m)=\sum_{k\in\ZZ } a_{k}(t-m)\ww x(k)
=\sum_{k\in\ZZ } a_{k}(t-m)x(k+m)=\sum_{k\in\ZZ } a_{d-m}(t-m)x(d).
\eaaa
This completes the proof of Corollary \ref{corrShift}.
 $\Box$

{\em Proof of Theorem   \ref{ThM}} follows immediately from Proposition \ref{propS1} and Corollary  \ref{corrShift}. $\Box$

 {\em Proof Corollary \ref{corrS1}}. 
 By  Theorem \ref{ThM}, the signal $x$ is uniquely defined 
 by the sequence $\{x(k)\}_{k\in\ZZ}$.  Further, it can be shown that the sequence $\{x(k)\}_{k\in\ZZ}$ represents  a  band-limited bounded discrete time signal as defined in Theorem 4 \cite{D23}.
Then Corollary \ref{corrS1}  follows from Theorem 4 \cite{D23}.
 $\Box$
 \section*{Concluding remarks}
\begin{enumerate}
\item
Since $|a_k(t)|\sim 1/k^2$ as $|k|\to +\infty$, we have that  $\sum_{k\in\ZZ}|a_k(t)|<+\infty$ for any $t\in\R$.
This allows to apply interpolation formula  (\ref{a}) to non-vanishing bounded signals. For these signals, the classical interpolation formula (\ref{Shan}) in not applicable, since the coefficients decay as $\sim 1/k$.
\item
It can be seen that selection of $N$ and $\O_1$ for  interpolation formula  (\ref{a0})-(\ref{a}) is non-unique.
Furthermore, it is possible that  there are other potential choices of $E$ in Proposition \ref{lemmaE}, leading to other versions of interpolation formula  (\ref{a0}).
\item
 The condition  that $\O\in (0,\pi)$, and that the  sampling points are integers, can be removed, as usual,  by linear changes of the times scale, i.e., with the replacement of  the signal $x(t)$ by  signal $x(\mu t)$, with $\mu>0$. Clearly, less frequent sampling would require $\mu>1$,
and  selection of a larger $\O$  would require $\mu<1$.
\item
The classical Whittaker-Shannon-Kotelnikov   interpolation formula (\ref{Shan})  allows spectrum bandwidth    $[-\pi,\pi]$. On the other hand, Theorem \ref{ThM} requires that the spectrum bandwidth  of $x$ is  $[-\O,\O]$, for  $\O\in(0,\pi)$.  Therefore, the possibility to cover non-vanishing signals is achieved via certain oversampling; this oversampling, however, can be arbitrarily small, since $\O$ can be arbitrarily close to $\pi$.
\item
It can be emphasised that the interpolation formula (\ref{a}) is exact; it is not an approximation.
Therefore, for a vanishing signal $x\in L_2(\R)\cap \V(\O)\cap C(\R)$, both formulae (\ref{Shan}) and (\ref{a})
give the same value. Similarly, for $x\in\V(\O_0)$ for $\O_0\in(0,\pi)$,  for all possible different choices of 
$\O\in[\O_0,\pi)$, $\O_1\in[\O,\pi)$, 
 $N=N(\O_1)$,  and  $E$,  the value of the sum (\ref{a0}) is the same. 
 Of course, the values  for the corresponding finite truncated sums will be different. 
 \item It is known that band-limited signals from $L_2(\R)$ are continuous. However, Theorem \ref{ThM}
is formulated for signals from $C(\R)\cap \V(\O)$ rather than for signals from $\V(\O)$, since  it is unclear yet  if  general type band-limited signals $x\in\V(\O)$ are continuous.  We leave it for the future research. 
\end{enumerate}

\end{document}